\documentstyle[aas2pp4]{article}

\lefthead{G\'{o}mez et al.}
\righthead{The Effects of Head-on Galaxy Cluster Mergers on Cluster Cooling Flows or Do Cluster Cooling Flows Survive Head-on Galaxy Cluster Mergers}

\begin{document}

\title{Do Cooling Flows Survive Cluster Mergers?}

\author{P. L. G\'{o}mez\altaffilmark{1}}
\affil{ Dept. of Physics \& Astronomy, Rutgers The State University
of New Jersey,\\ 136 Frelinghuysen Road, Piscataway NJ 08854-8019,\\
 Email: percy@physics.rutgers.edu}

\author{C. Loken\altaffilmark{1,2}, K. Roettiger, J. O. Burns\altaffilmark{1}}
\affil{Dept. of Physics \& Astronomy, University of 
Missouri - Columbia, \\Columbia MO 65211,\\
 Email: cloken@ap.stmarys.ca,
kroett@hades.physics.missouri.edu, burnsj@missouri.edu}

\slugcomment{Submitted to the Astrophysical Journal}

\altaffiltext{1}{Much of the work reported here was performed while at the Department
of Astronomy, New Mexico State University, Las Cruces, NM} 
\altaffiltext{2}{present address: Dept. of Astronomy \& Physics, Saint Mary's University,\\ Halifax, NS,
B3H 3C3, Canada}



\begin{abstract}

We report the results of recent numerical simulations 
of the head-on merger of a cooling flow cluster with an infalling
subcluster of galaxies. The objective of these simulations was to examine the 
effects of different types of cluster mergers (with 16:1 and 4:1 mass ratios) 
on the evolution of cluster cooling flows (with mass accretion rates of 100 
and 400 M$_{\odot}$/year). The 2-dimensional simulations were performed 
with a combined Hydrodynamics/N-body code on a uniform grid  with a 
resolution of 20 kpc ($\sim$ 12 zones/core radius).

In our simulations, cooling flow disruption is indicated
by a dramatic increase (by a factor of 10-40) in the central 
cooling time of the primary cluster.
We find that the ram-pressure of the infalling gas is crucial in determining 
the fate of the cooling flow as disruption occurs when a substantial
amount of subcluster gas reaches the primary's core. In such cases, the 
subcluster gas can increase the central cooling time by displacing the 
high-density 
cooling gas and by heating it via shocks and 
turbulent gas motions.   
However, the fate of a merging cooling flow is also dependent on its 
initial cooling time. In cases where the initial cooling time is 
very small (i.e., 10-40 times smaller than the Hubble time) then, even
if the flow is disrupted, the central cooling time will remain less 
than a Hubble time and the flow will likely re-establish itself. 
This has an important observational
consequence because such clusters will be classified as cooling flows
on the basis of their cooling times
even though they have experienced a significant merger.
In addition, we find that there is a time delay between core-crossing
and the point at which the central cooling time of a disrupted flow becomes 
of order a Hubble time. Thus, even in the case of disruption, a cluster
can be classified as a cooling flow and exhibit substructure (indicative
of a merger) for 1-2 Gyr after merging with a subcluster. 
We argue that our results make it possible to reconcile the
high cooling flow frequency inferred by some observations with both
high merger rates and a high frequency of substructure.

\end{abstract}




%

\section{INTRODUCTION}

Clusters of galaxies form at the intersection of sheets and filaments
in the ``Cosmic Web'' through a process of accretion and 
mergers with other clusters (e.g., Bond, Kofman, \& Pogosyan 1996;
Colberg et al.~1999). Thus, mergers are very likely a generic element
of cluster evolution. Their effects on cluster morphologies and 
properties (e.g., Burns 1998) have been investigated via 
high-resolution numerical 
simulations which reproduce many of the features in observed 
merger candidates (e.g.,  Evrard et al. 1993; Roettiger, Burns \& Loken 1996, Roettiger, Stone \& Mushotzky 1998). Numerical
simulations are essential for understanding the details of cluster physics 
and evolution as well as for interpreting observations which attempt
to use clusters in order to constrain cosmological parameters (e.g.,
Gunn 1978; Birkinshaw 1979; Richstone et al.~1992; Mohr et al. 1995).

Another common feature of cluster evolution appears to be the formation of cooling
flows (e.g., review by Fabian 1994). A large fraction ($> 50$\% of x-ray bright clusters; e.g., Edge, 
Stewart, \& Fabian 1992;
White, Jones, \& Forman 1997; Edge 1997) of massive clusters are believed to harbor
cooling flows which may be the natural outcome of cluster evolution
over a wide range in cluster masses (Knight
\& Ponman 1997). A cooling flow evolves in a cluster atmosphere because the X-ray emission is a very effective cooling mechanism. 
The typical physical properties of the gas located at the core of a cooling 
flow cluster (i.e., n $\sim$ 0.01 cm$^{-3}$ and  T 
$ \sim 10^7 K$) imply a cooling time less than the age of the Universe. 
Therefore, gas  surrounding this core, which cannot cool as fast as the 
central gas, loses pressure support and starts a subsonic inward flow. 
Note  that most of the evidence supporting the presence of cooling flows 
in galaxy clusters is indirect because current X-ray telescopes lack the
spectral resolution to directly measure velocities via emission lines. 
Thus, cooling flows are inferred to exist from observed cluster gas 
temperature gradients (e.g., White \& Silk 1980; Allen et al.~1995), 
from the detection of excess emission in the cluster core 
(e.g., Stewart et al.~1984; Allen et al.~1995), or from the presence of $\sim 10^4$K gas in cluster cores as revealed by 
observations  of H$\alpha$ filaments (Heckman et al. ~1989; 
Donahue et al.~1992; Crawford et al.~1995).

Is there any correlation between the presence of cooling flows and
the  presumed merger history of a cluster? Donahue et al.~(1992) found 
evidence of more cooling flow clusters in the past than in the present by 
studying their optical H$\alpha$ emission. In their analysis, the percentage 
of cooling flow clusters increases to 56\% in the redshift range 
0.2 $<$ z $<$ 0.37 from 35\% in the 0.065 $<$ z $<$ 0.14 range. 
This result can be explained if mergers are more frequent in the present than 
in the past {\it and} if cluster mergers disrupt cooling flows. However,
this putative increase in cooling flow frequency with redshift is 
contradicted by Crawford et al.~(1995) who analyzed 
a sample of 20 clusters with z $>$ 0.2 and concluded that only 35\% of them 
show evidence of H$\alpha$ emission.
To complicate matters even more, the suggestion that there are relatively
few nearby cooling flow clusters has been questioned since some observations 
suggest that as many as (70-90)\% of bright nearby clusters have cooling 
flows (e.g., Edge, Stewart, \& Fabian  1992; White, Jones, \& Forman 1997; 
Peres et al.~1998) with 39\% of them showing optical line emission
(Edge \& Stewart 1991). 

Detailed multiwavelength observations of some galaxy clusters seem to 
support the hypothesis that cluster mergers affect and in some cases even
destroy cluster cooling flows. The presence of X-ray and/or optical 
substructure is used to infer a dynamically complex system and/or a 
recent cluster merger where the gas is not in hydrostatic 
equilibrium with the gravitational potential (Davis \& Mushotzky 1993, White et al. 1993, Mohr et al. 1993). The Coma cluster (e.g., Burns et al.~1994), 
A2256 (e.g., Briel \& Henry 1994; Roettiger, Burns, \& Pinkney 1995; 
Markevitch 1996), A2255 (Burns et al.~1995) and A3627 
(e.g., B\"{o}hringer et al.~1996) are some 
examples of massive clusters that appear to have undergone a recent merger 
but do not have cooling flows. On the other hand, A1664 (Allen et al.~1996), 
A2597 (Sarazin et al.~1995, G\'{o}mez et al., in preparation),  A2390 and A2667 (Rizza 
et al.~1998), and A2065 (Markevitch et al.~1999) are examples of clusters 
with evidence for both a cooling flow {\it and} x-ray substructure. 
All these results may make sense in view of the 
relation between cluster substructure and cluster cooling flow 
strength uncovered by Buote \& Tsai (1996) in a study of 23 clusters.
They quantified the dynamical state of a cluster and its substructure by 
using a power ratio technique that consists of expanding the cluster X-ray
surface brightness into gravitational multipoles. In this manner, they find 
that the ratio of quadrupole-to-monopole (a measure of cluster substructure)  
decreases as the cooling flow strength increases. 
Their findings are consistent with a scenario in which recent mergers 
can either diminish the  strength of a cooling flow (e.g., A1664) or even 
destroy them (e.g., Coma, A2256). If we assume that the substructure was 
created by a recent cluster merger (and the substructure has a lifetime
short compared to the time necessary to re-establish a disrupted cooling flow), 
then  mergers may not always destroy the cluster cooling flow.

Despite the apparent ubiquity of both the cooling
flow phenomenon and mergers, no detailed numerical studies of the effects of mergers on cooling 
flows have been performed to date. 
Early N-body simulations with no gaseous component (McGlynn \& Fabian 1984) 
and more recent analytic studies (Fabian \& Daines 1991) have suggested that 
mergers destroy cooling flows but these
studies have limitations (for example, Fabian \& Daines considered only
subsonic mergers between clusters of roughly equal mass) that prevent
applying their results to the wide range of parameters (masses, cooling
flow strengths etc.~) which characterize actual mergers. For instance, the typical cluster gas sound speed 
is $\sim$ 1,000 km s$^{-1}$ while the free-fall relative velocity between the 
merging clusters could be as high as 3,000 km s$^{-1}$. These estimates are consistent with the values derived by 
Markevitch et al.~(1998b) in Cygnus A, A3667, and A2065. Therefore, this 
is a complex problem that needs the more sophisticated treatment provided 
by numerical simulations.

In order to study the effects that cluster mergers have on cooling
flows, we have performed a set of new 3D N-body with 2D Hydro numerical simulations 
(assuming cylindrical symmetry and with 20 kpc resolution)
of head-on cluster mergers in which we follow the evolution of a cooling flow located in the most massive cluster (i.e., the primary cluster). With this study, we propose to answer the following questions:  (1) What type of merger affects a cluster cooling flow? and (2) If a cooling flow is disrupted (i.e., the core is heated and the cooling time increases), what are the 
merger parameters that determine if a new cooling flow can be formed?

We have organized this paper as follows. In \S2, we describe the code used in the simulations, the initial conditions,  as well as some of the code tests. Section 3 presents the general results from all the simulations. Next, we discuss the possible interpretations of the numerical simulations in \S4. Finally, we summarize our conclusions in \S5. We use H$_o$=75 km/s/Mpc and q$_o$=0.5 throughout the paper.

\section{NUMERICAL METHOD}

All the simulations presented here were performed with a hybrid code
similar to the one used by Roettiger et al.~(1993, 1996, 1997) that combines 
the N-body code TREECODE (Hernquist 1987; with a softening parameter of 0.2 
and a tolerance of 0.7) with the Eulerian, finite-difference fluid 
dynamics code ZEUS-3D (Clarke 1990; Stone \& Norman 1992).
The main differences between our code and Roettiger et al.'s code are that (1) we increased the resolution by a factor of 2.5 by performing these simulations in 2 dimensions and invoking cylindrical symmetry along the merger axis, (2) we used a different Poisson solver for computing the gravitational potential from the N-body particles, and (3) we included radiative cooling.

The only link between the hydro and N-body codes is through the 
Poisson solver. Every time an N-body step is required, the evolving 3-D 
N-body particles are rebinned in 2-D. Next, we determine the boundary 
conditions by computing the contribution of each particle at every boundary
cell. Finally, we solved the finite-differenced Poisson equation in 2-D by 
using the Generalized Conjugate Residual method (GCR; Eisenstat et al.~1983) within the NonSymmetric PreCondition Gradient package (NSPCG; Oppe et 
al.~1988). Typically, there are 5-7 hydro time steps (governed by the Courant condition) for every N-body step. Our implementation of the code does not include the self-gravity of the gas which
is not expected to be important since the gas component comprises less than $\sim$15\% of the 
total dynamical mass.

The cooling function that we used is an analytical approximation 
to the cooling curve (Raymond \& Smith 1977) based on the equations
given by Westbury \& Henriksen (1992) for the case of half-solar 
abundance. Our modeling of the effects of radiative cooling is 
computationally expensive because it includes the solution of the 
energy equation in every cell at every time step. We avoid the problem 
of catastrophic cooling at the cluster core with the use of a mass 
drop-out term intended to simulate the mass loss produced by star formation. 
This term follows the prescription of Sarazin and White (1985) for mass 
loss ($d \rho / d t = q \rho /t_c$, where q is the mass drop out term 
and $t_c$ is the cooling time). Unfortunately, this extra term forces us 
to solve the continuity and energy equations as two coupled implicit 
equations by using the Newton-Raphson method.  However, the use of 2D 
cylindrical symmetry allows us to perform these simulations with significant 
savings in the run time. Note that the amount of gas that drops out during 
the entire simulation is gravitationally negligible (less than 2.5$\%$ of 
the total core mass).

In order to follow the evolution of the primary and secondary cluster gas 
during the merger, we have added two passive (i.e., dynamically insignificant)
scalars (or tracers) to the code. Each cluster was initialized with a passive scalar distribution that mimicked its initial cluster gas distribution. As the simulation progressed, the passive scalars were advected by the velocity field. In this way, we were able to trace the motion of each cluster's gas during the merger. 

In order to test and verify our new code, we performed the same cluster 
tests used by Roettiger et al.~(1997) which consisted of evolving a cluster
composed of only N-body particles, N-body particles and gas, and the uniform
motion of an isothermal cluster composed of N-body particles
and gas across the grid. We found excellent
agreement between these tests. We also verified that we could move
an isolated cluster with a steady-state cooling flow across the grid at 600 
km/s with no major changes in the cooling time, the central
temperature, or in the radial density profile. This test is essential
to our demonstration (\S3.3) that some mergers are able to disrupt a 
cooling flow.

\subsection {Parameter Space}
	
McGlynn and Fabian (1984) suggested that a merger of two similar 
clusters would destroy a cluster cooling flow. Indeed, recent numerical simulations (most without radiative cooling and none with cooling flows) performed by various groups (e.g., 
Roettiger et al.~1993, ~1996; Schindler \& M\"{u}ller 1993;
Pearce, Thomas, \& Couchman 1994) have shown that 
these massive mergers have profound effects on the properties of the cluster gas. For instance, these mergers generate large gas bulk flow motions and turbulence, especially, within the core. Thus, we decided to study the effects of mergers with lower 
mass subclusters since we assumed that mergers between similar clusters 
(mass ratio of 1:1 or 2:1) would very likely destroy cooling flows. 

The first parameter that we explored was the total mass of the subcluster. 
The sudden inflow of the subcluster will cause a rapid fluctuation of the
gravitational potential that could affect the cluster cooling flow region. 
For our simulations, we have chosen a primary cluster with a dark 
matter mass of 10$^{15}$ M$_\odot$ (i.e., Coma-like) and subclusters with 
1/16 (0.625 x 10$^{14}$ M$_\odot$) and 1/4 (2.5 x 10$^{14}$ M$_\odot$) 
the primary cluster mass (see Table 1). 

The amount of gas in the infalling secondary cluster is also likely to be
an important parameter. Increasing the baryon fraction in the secondary
cluster will increase the ram-pressure of the infalling gas which,
if it reaches the primary core, may have significant dynamic or thermal 
effects on the cooling flow. Thus, we address the question of whether
gas-rich subclusters are more likely 
to affect the primary cooling flow than gas-poor subclusters by varying
the gas mass fraction in the subcluster from 1.2\% to 15\%. We did not attempt to go higher than 15\% because our code does not include gas self-gravity. Note that poor clusters typically have low gas fractions ($\sim$ 5-30\% Mulchaey et al. 1996).

The last parameter varied was the strength of the cooling flow 
(i.e., \.{M} or the mass accretion rate). We chose two primary cluster cooling strengths: 100 and 400 M$_\odot/$year. We hypothesized that a stronger cooling flow should be more difficult to destroy by a merger than a weak cooling flow. This is because a stronger cooling flow would have a very steep central density profile, and thus, it would be more likely to dissipate the effects of the shock formed during the merger. Table 1 shows the parameter combinations used in our simulations.

\subsection {Initial Conditions}

The initial conditions for the simulations were simple and idealized. We started with a massive primary cluster with a cooling flow and an isothermal secondary cluster that merge under the influence of their mutual gravity.
They were placed on a cylindrical grid with dimensions 500 x 150 zones 
(assuming azimuthal symmetry) corresponding to a resolution of 20 kpc per 
zone ($\sim$12 zones across the primary cluster core radius) and given an 
initial relative velocity of 600 km/s which is consistent with the mean
peculiar velocity of nearby galaxy clusters (Colless et al. 1999). The 
initial separation between the clusters was 4.9 Mpc for the 1:4 mass 
ratio mergers and 4.4 Mpc for the 1:16 mass ratio mergers. This setup has
two advantages. First, it puts the clusters far enough apart so that they do 
not severely affect each other. Second, the clusters are close enough
so that the merger occurs fairly quickly, thus, saving computational time.

	There are several advantages in using these simplified initial conditions (e.g., Roettiger et al.~1993; ~1997, Pearce et al.~1994) over cosmological initial conditions. First, the use of already formed clusters allows us to use a relatively small grid, thus, enhancing the spatial resolution. Second, the symmetry of the set-up allows us to perform the gas simulations in 2D which saves run time and memory requirements. Finally, we have a very well defined baseline with which to compare the subsequent evolution making it
straightforward to determine how the cluster properties are affected by the merger.

   The collisionless dark matter was represented by N-body particles distributed spatially according to a lowered isothermal King model (King 1966) characterized by a concentration parameter of 1.08 (Binney \& Tremaine 1987). The primary cluster dark matter in each simulation was represented by 30,000 N-body particles whereas the number of particles in the secondary clusters was scaled accordingly so that each N-body particle has the same mass. Table 1 includes the number of particles per subcluster.

	The secondary cluster gas was initially isothermal and in hydrostatic equilibrium with its dark matter gravitational potential. Thus, the shape of the gas distribution (we assume $\beta =1.0$) and the gas temperature were obtained by solving the equation of hydrostatic equilibrium. This approach leaves the central density of the secondary cluster as a free parameter. However, our choice for the central density was limited by the fact that we do not include self-gravity in the code; thus, we chose an overall gas fraction $\leq$ 15\%. This choice also assures us that the central cooling time for the subcluster is much larger than a Hubble time. Other cluster parameters appear in Table 1.
	
The cooling flow cluster was assembled in a different manner. First, 
we computed the gravitational potential produced by the primary cluster 
dark matter distribution. Next, we laid down an isothermal gas distribution 
in hydrostatic equilibrium with this gravitational potential onto the 
ZEUS grid. Then, we turned on cooling and followed the evolution of the 
cluster core until it reached a steady-state cooling flow (this process 
normally takes $\sim$ 5-7 cooling times). Table 2 shows the initial parameters 
of the pre-cooling flow isothermal clusters used in the cooling flow model 
evolution. In order to avoid the problem of catastrophic cooling at the 
cluster core and as a way to represent the multiphase nature of the 
cooling gas, we used a mass drop-out value of 0.2 in the simulations 
($q_{drop}=0.2$; Westbury \& Henriksen 1992). Finally, we extracted 
the 1D density, temperature, and radial velocity  profiles and used 
them as the initial conditions for the primary cluster cooling flow. 
The two steady-state cooling flows were measured to have mass accretion 
rates (\.{M}) of 100 and 400 M$_\odot$/year at the cooling radius (where 
the cooling time is equal to the age of the Universe). Figure 1 shows the 
density and temperature profile of our cooling flow models. Note that 
the central region of the cluster ($<$ 100 kpc) has a lower temperature 
and a higher gas density than the surrounding cluster gas.

\section {RESULTS}

\subsection {Dark Matter Evolution}

As has been pointed out before (e.g., Roettiger et al.~1997), the most 
important effect due to the dark matter evolution is the sudden fluctuation of the gravitational potential minimum during core crossing. Figure 2, which shows the evolution of the gravitational potential minimum as a function of the time, confirms that the most dramatic change occurs during core crossing in the 4:1 mass ratio cluster merger. At this time, the gravitational potential suffers a sudden and relatively brief ($\sim$ 1 Gyr in duration) deepening caused by the subcluster. Moreover, the oscillations in the gravitational potential, which are caused by the secondary cluster falling back into the primary cluster, suggest that the secondary cluster dark matter
distribution survives the first pass through the primary cluster core. On the other hand, the lack of gravitational potential oscillations and the presence of a spray of N-body particles exiting the primary cluster points toward the destruction of the secondary cluster during the 16:1 merger. Note that the most dramatic changes in the gravitational potential, which should cause mixing and heating of the gas, occur in the 4:1 mass ratio merger. Are these changes strong enough to affect the gas properties of the cooling flow region? We will address this question in \S4.

\subsection {Cluster Gas Evolution}

In general, the evolution of the cluster gas in our nine merger simulations follows the same patterns described in the analysis of other numerical simulations (e.g., Roettiger et al.~1993, 1997; Schindler \& M\"{u}ller 1993; Pearce et al.~1994). To facilitate our explanation of the effects of mergers on 
cooling flows, we will concentrate on the gas evolution in two simulations: 
\#7 and \#8 (Table 1). 
These simulations  are both  16:1 mass ratio mergers involving
identical cooling flow clusters and differ only in the  the secondary 
cluster's total gas content. We chose these two examples because the 
cooling flow (e.g., central region of low temperature and high density) 
survives the effects of the merger \#7 while it is disrupted in 
merger \#8. 

Figure 3 consists of 6 contour plots showing the evolution 
of the logarithm of the gas density in merger \#8 (Table 1). The density contours are overlaid onto a grey scale plot that represents the distribution of the secondary cluster passive scalar (\S2) which traces the 
secondary cluster gas. Note that the times are relative to the core passage and that in all the panels the contours and grey levels are scaled to the same values. In the first two panels, we see the subcluster falling towards the 
primary cluster from the right and creating a bimodality in the gas 
density distribution. The leading
edge of the subcluster is compressed and develops into a bow shock as
the subcluster's motion soon becomes supersonic (at $\sim$ 0.6 Gyrs before core crossing). 

The morphological effects of the merger are most strongly evident at the 
time of core crossing and shortly thereafter. For instance, the primary
cluster core suffers several changes as it evolves from being spherical 
before core-crossing to an elliptical shape at t=0. Later, the core 
exhibits extreme isophotal twisting (t=0.25 Gyrs) which relaxes
back into an elliptical core (t=1 Gyrs). Eventually, the merger remnant
will re-adopt a more circular shape (t $\sim$ 5.5 Gyrs). The most
interesting morphology occurs at t=0.25 Gyrs when the core (inner $\sim200$ kpc) of the cluster shows two
distinct elongations in the density distribution. One elongation is parallel 
to the merger axis and is likely caused by the reaction of the cluster gas 
to an elongated gravitational potential which is aligned along the merger axis. 
This elongation forms slightly before the time of core crossing and lasts 
at least 1 Gyr. The second elongation is perpendicular to the merger axis, 
lasts for 0.75 Gyrs, and is 500 kpc wide. The grey scale plot suggests that 
this elongation is caused by compressed secondary cluster gas that is 
mixing with and impinging on primary cluster gas.

Interestingly, the bow shock that appears at the leading edge of the secondary cluster gas
distribution protects the subcluster gas from significant mixing with 
the primary cluster gas until t=0.25 Gyrs. Note that the secondary gas
was effectively stripped from its potential earlier (the subcluster DM 
passes through the primary cluster core at 0 Gyr).
However, at t=1 Gyr, secondary cluster gas has started to penetrate the 
primary cluster core and by t=5.5 Gyrs, gas from the secondary cluster is readily mixing with primary cluster gas as can be seen in Figure 4. This figure  shows line plots (along the merger axis) of the total gas density and the subcluster passive scalar at three different epochs.

Figures 5 depicts the same quantities and epochs as Figure 3 but for 
merger \#7. There are two main differences in the gas evolution between these two mergers. First, the bimodality of
the pre-merger gas distribution in merger  
\#7 is not as evident as in merger \#8. 
This is simply a reflection of the fact that the secondary cluster
gas distribution is not as dense in \#7 as it was in  merger \#8. 
Second, subcluster gas penetrates more deeply into the primary cluster core,
and in greater amount, in merger \#8 (see Figure 4).
Note that at t=5.5 Gyrs, secondary cluster gas is present at 
the core of the primary cluster in both simulations; however, the density 
of the secondary cluster gas located within the cooling flow core is greater 
in merger \#8 than in \#7 as seen in Figure 4 (by almost a factor of 2).

Figures 3 and 5 also allow us to track the evolution of the cooling flow
through the mergers. We can identify the cooling flow region as the very dense knot located at the primary cluster core. Note that at t=5.5 Gyrs, 
the primary cluster core in the gas rich secondary merger (\#8, Figure 3) 
is less dense than it was before the merger (Figure 4). This demonstrates 
that the cooling 
flow properties have been affected by the merger. Moreover, the core of this cluster has expanded significantly due to heating and
non-thermal pressure support (i.e., turbulence).
This can be seen by analyzing Figure 4 and by comparing the evolution of the primary cluster core as depicted in the different panels in Figure 3. On the other hand, the primary cluster central
density appears to remain unaffected throughout merger \#7 (Figure 5).

\subsection {Cooling Flow and Gas Temperature Evolution}

The contour plots of the gas density evolution suggest that the properties 
of the cooling flow region were
more strongly influenced by the high gas density subcluster (merger \#8) 
than by the low gas density subcluster (merger \#7). In order to expand this analysis and determine how the merger has affected the gas temperature, we present a comparison of the temperature evolution for these two mergers.

Figure 6 shows grey scale plots of the evolution of the spatial distribution 
of the gas temperature for three epochs of simulations \#7 and \#8. Our 
analysis reveals a number of interesting features.
First, there is evidence of the heating created by the bow shock that 
develops at the boundary between the two subclusters. We have examined this 
region of the cluster and determined that the strongest shock first 
appears at $\sim$ 500 kpc from the primary cluster core and is 
generated by gas moving at peak Mach numbers $\sim$ 1-2. 
Second, this shock penetrates deeper into the primary 
cluster for merger \#8 (gas rich subcluster) than for merger \#7. This 
suggests that the degree of penetration of the secondary cluster 
gas into the primary cluster depends on its relative momentum.  We will 
discuss this possibility in more detail in the next section.
Third, the cooling flow can be identified as the very cold (dark) region 
located at the center of the primary cluster in the first panel of these 
two figures. 
This cool region has all but disappeared at t=5.5 Gyrs in merger \#8. 
The fact that the peak density in the primary cluster core has decreased 
while the minimum temperature has increased indicates that the cooling
flow has been disrupted in this merger. Simulation \#7 shows no such
signatures and therefore the cooling flow has not been disrupted in this
case.

This qualitative evidence for the disruption of the cooling flow can
be quantified by reference to Figure 7 which shows the evolution of
the cooling time in the primary cluster core. This figure clearly
reveals the fate of the various cooling flow mergers we investigated.
Mergers such as \#8 show a dramatic increase in the cooling time
after the merger while there is no change in the central cooling time
for run \#7.
Since $t_{cool} \propto T^{1/2}/\rho$, any significant increase of 
the cooling time is an indication of an increase of the gas temperature 
and/or a decrease of the gas density. We have already noted that the
central gas density decreases significantly in the case of disruption
(run \#8, see Figure 4). To assess the role of the temperature, we
plot the evolution of the central gas temperature in Figure 8.
The temperature evolution is remarkably similar to that of the
cooling time but we note that the temperature change does not account
for the entire change in $t_{cool}$. In fact, for merger \#8, the
density decrease (Figure 4) is more significant than the temperature increase
in lowering the cooling time. Of course the two effects are closely related
as heating of the core will result in expansion and a lowering of the 
central gas density. 

What are the crucial parameters which determine whether a merger
will disrupt a cooling flow?  The only clear trend is that the likelihood
of disruption increases as the amount of gas in the secondary cluster
increases. This behavior can be seen by considering mergers \#1, \#2,
and \#3 which differ only in the central density of the secondary cluster.
There is strong disruption in the case where the secondary's central density
is highest and only mild disruption in the others (see Fig. 7 and Table 1).
The same trend can be seen in the case of a different mass ratio
(\#6 and \#5) and in the case of a stronger cooling flow (\#8, \#7 and
\#9). It is difficult to discern, or even isolate, any systematic trends
with cooling flow mass accretion rate or the mass ratio of the two clusters.

We find two interesting results which have consequences for 
reconciling observations of  high cooling flow frequencies with
high merger rates. First, we observe a time delay of typically
1-2 Gyrs between the time of core crossing and the time at which the 
cooling flow is disrupted (consider, e.g., mergers \#6 and \#4
in Fig. 7). Thus, even when the cooling flow will be strongly disrupted
(e.g., merger \#1), an observer could detect the signatures of substructure 
and a central cooling flow so long as the central cooling time is less
than a Hubble time, i.e.~for up to 2 Gyrs after the merger has taken place.
As we will discuss later, this fact also supports the contention that
gas dynamics (which will lag behind the N-body dynamics) accounts for
the disruption of the cooling flow. Secondly, we point out that the 
fate of a cooling flow is dependent on its initial cooling time. In both 
mergers \#6 and \#8, the central cooling time rises by a
factor of $>20$ within 3 Gyrs of core-crossing indicating that the initial
cooling flow has been disrupted. However, since the initial cooling time
for merger \#8 was very small ($\sim 0.1$Gyr), its final cooling time
is still significantly less than a Hubble time and it would be 
observationally classified as a cooling flow. Thus, cooling flows with very short cooling times ($\leq
0.2$ Gyr), can be significantly affected by a merger yet still appear
to be cooling flows.  Furthermore, because the final cooling time is short, 
the original flow is likely to be quickly re-established in these
cases.

\section {DISCUSSION}

\subsection {What Destroys the Cooling Flow?}

In the previous section, we showed that some types of cluster mergers 
are able to destroy cooling flows while other mergers will leave them intact. 
In this section, we will review some of the possible mechanisms triggered by 
a cluster merger that can affect the gas properties of the cooling flows. 

	One of these mechanisms is the violent increase in the depth of the gravitational potential at the time of core crossing. However, our analysis of
the simulations indicates that this process is not the most important
factor in determining the future of the cooling flow. For instance, we have shown that identical mass ratio mergers affect the cooling flow in different ways since we find a wide variety of outcomes in the 16:1 mass ratio mergers. Thus, there are no effects on the cooling flow produced by mergers \#7 and \#9 while the cooling flow does not survive merger \#6. Furthermore, there is a delay between the moment of core crossing and when the cooling flow starts to feel the merger effects in our simulation (e.g., $\sim 1.5$ Gyrs in the
case of merger \#8). This delay suggests that the gravitational potential increase at the time of core crossing is not enough to destroy the cooling flow. Finally, in an effort to better isolate the effects of the gravitational potential from the effects caused by gas dynamics, we ran a  16:1 mass ratio merger (\#9) of a cooling flow cluster with an essentially gas-free secondary cluster. We note that this merger does not significantly affect the 
gas properties of the cooling flow region. The different reaction of the 
cooling flow to the same mass ratio mergers is an indication that
the most important mechanism responsible for disrupting the cooling flow
is the interaction of primary and secondary cluster gas.

Another potential mechanism responsible for the disruption of
the cooling flow region is shock heating of the gas. A 
shock is generated by the supersonic infall of the secondary cluster gas into the primary cluster core. For instance, the infalling secondary gas develops a Mach 2 shock when its leading edge is located at $\sim$ 200$-$500 kpc from the
primary cluster (e.g., merger \#7). When the clusters are separated by less
than 200 kpc, it is
difficult to identify the shock structure due to the steep density profile near
the core and in the cooling flow. Moreover, the shock decelerates as it
moves closer to the core as it encounters an increase in the ambient density.
Even if this Mach 2 shock were to penetrate all the way to the core of the primary cluster, it would only increase the temperature by a factor of 
$\sim$ 2.1. This limited amount of heating could not cause the large central
temperature jumps observed in mergers 
\#1, \#4, \#6, and \#8. However, this heating could explain the small 
disruption caused in other mergers (e.g., \#3).

Our results support Fabian \& Daines (1991) suggestion that the most important factor in determining whether a cooling flow can survive a merger is the ram pressure of the infalling secondary cluster. This could explain why mergers that differ only in the secondary cluster gas content have different effects on 
cooling flow evolution. In order to test this idea, we have compared the relative infall velocity of the secondary cluster ($v_{s}$) with a threshold velocity ($v_{bal}$). We define $v_{bal}$ as the velocity that balances the cooling flow thermal pressure with the ram pressure produced by 
the motion of the subcluster gas (i.e., $v_{bal}^2 = P_{CF}/\rho_{sec}$, where $P_{CF}$ is the primary cluster thermal pressure, and $\rho_{sec}$ is the secondary cluster central density). The right-hand panel of Figure 9 shows a plot of $v_{bal}$ as a function of radius for all of our 4:1 simulations while the left-hand panel shows the same plot for the 16:1 mass ratio mergers. Furthermore, we have overlaid on these plots (thick line) our estimate for the
$v_{s}$ of the secondary cluster. This estimate was computed from the N-body particles and it is expected to be an upper limit for the actual infall velocity of the secondary cluster gas which decreases as it is stripped from its
gravitational potential. Moreover, in order to compute $v_{bal}$, we have assumed that the secondary cluster peak density remains constant throughout the merger. If we keep these considerations in mind, we note that there is
a  trend for cooling flows to survive in cases where $v_{s} \le v_{bal}$
within some radius. One exception appears to be \#4 which is classified as
a strong disruption (Table 1) but we note that even in this case, the
final cooling time is less than 2 Gyrs which suggests the flow will quickly
re-establish.
Therefore, this plot suggests that substantial amounts of secondary cluster 
gas manage to penetrate into the cluster cooling core only when the 
ram pressure is larger than the local cooling flow thermal pressure. 
Note that this ram pressure model is consistent with our results. It also
explains naturally why mergers of identical cooling flow clusters and
subclusters that differ only in their overall dark matter mass ratio 
can have different effects on the cooling flow (\#4 and \#7). The subcluster 
in the 4:1 mass ratio merger has a greater momentum than the subcluster 
in the 16:1 merger (since the 4:1 mergers have a larger infall velocity 
and core radius) and its gas is therefore able to penetrate and disrupt
the primary cooling flow.

Finally, the time-delay (1-2 Gyr) between core crossing and cooling flow
disruption further supports the idea that gas processes are ultimately
responsible. Previous numerical simulations (e.g., Roettiger et al. 1998; 1999)
have shown a 1-2 Gyr delay between core passage and the onset
of turbulence which can heat the gas core as it supplies a  non-thermal component
of pressure support equivalent to $\sim$20\% of the core's thermal pressure.

\subsection{Later Stages of the Cooling Flow}

Will the disrupted cooling flows  re-establish themselves? 
Figure 7 demonstrates that the initial cooling time is critical. For example,
mergers \#4 and \#8 resulted in a relative increase in central 
cooling time similar to 
that experienced in mergers \#1 and \#6 but their cooling times never
reach a Hubble time and they even drop a few Gyr by the end of the
simulations. Thus only the flows with very short initial cooling times
can re-establish themselves after experiencing a significant merger.
The severity of the merger also plays a role. For example, merger \#1 resulted
in disruption and a final central cooling time of $\sim$30 Gyr while the
identical merger involving lower baryon fraction subclusters
(\# 2 and \#3) caused a relatively mild increase in the cooling time.
These mildly affected clusters may also re-establish themselves.

\subsection {Radio Sources, Cooling Flows, and Cluster Mergers}

Abell 2597 is an interesting cooling flow cluster (cooling rate $\sim$ 327 M$_\odot$/year, Sarazin et al.~1995) which shows some evidence of X-ray and optical substructure (Sarazin et al.~1995, G\'{o}mez et al., in preparation) and the presence of a disrupted compact (size $\sim$ 20 kpc) tailed radio source located at the {\it cluster core} (PKS 2322-122, Owen et al.~1992). Sarazin et al.~have examined in detail the radio structure of PKS 2322-122 and determined that one of its jets is sharply bent by more than 90$^0$. In their analysis, they propose several models for the jet bending. One of those models suggests that the bending is caused by the interaction between the jet and a large transversal bulk flow of cluster gas (with velocities $\sim$ 1000 km/s) fueled by a recent cluster merger. Our simulations are not consistent with that dynamical scenario. We find that high velocity flows near the cluster center, such as those 
required to bend the jets, are inconsistent with the presence of a cooling
flow. The mergers that leave the cooling flow unaffected can
also produce turbulence and bulk flow gas motions but at much greater
distances ($>150$ kpc) from the cluster center.
Thus, we find that, given the limited spatial resolution of our simulations, large 
bulk flows ($\sim$ 1000 km/s) and cooling flows do not occur within the 
same region. Therefore, a merger could not maintain a significant cooling 
flow and produce the gas dynamics needed to bend an extended radio source
located at the cooling flow region.

\section {SUMMARY AND CONCLUSIONS}

We have performed a set of 9 simulations of head-on mergers of two galaxy
clusters that include, for the first time, the effects of radiative cooling. 
The objective of the simulations was to determine how different mergers 
would affect a cluster cooling flow. The parameter space 
covered by this study consisted of a range of mass ratios for the
two clusters (16:1 and 4:1), a range of overall baryon fractions of the secondary cluster (from 1.2\% to 15\%), and a range of strengths of the primary cluster cooling flow (100 and 400 M$_{\odot}$/year). The simulations were performed with a combined Hydrodynamics/N-body code on a grid with a 
resolution of 20 kpc. The N-body particles were evolved in 3-D while the 
gas evolution calculations were performed in 2-D because there is symmetry 
around the merger axis. 
 
We find that some cluster mergers destroy cooling flows while others leave them intact. Our analysis of these results lead us to draw the following three
conclusions. First, the destruction of the primary 
cluster cooling flow depends principally on the ram-pressure of the 
gas in the infalling cluster. In accordance with this, we found that 
if all other parameters are kept fixed, the likelihood of disrupting a 
cooling flow increases as the baryon fraction in the secondary 
cluster increases. We attribute the disruption to two factors linked to
the gas dynamics of the merger; the ram-pressure of the subcluster gas is 
able to displace the high-density gas in the cooling core as well as 
heating it through adiabatic compression, shocks and turbulence.
Second, the time-scale on which a new cooling flow re-establishes itself
depends on the initial cooling time of the cluster as well as on the 
severity of the merger. The post-merger cooling times of initially
short cooling-time flows are less than a Hubble time and thus they may
re-establish themselves quickly. Third we note that, in the case of 
disruption,  there is a lag of at least 1-2 Gyr between the epoch of the 
merger (when the dark matter cores cross) and the point at which the 
central cooling time increases to the Hubble time. These last two 
conclusions have important consequences for interpreting observational
determinations of the frequency of cooling flow and substructure occurrence.
Any cooling flow cluster involved in a merger will exhibit some degree of
substructure {\it and} appear to have a cooling flow for at least 1-2 Gyrs
after core-passage. Furthermore, if the initial cooling time is very 
short (10-40 times less than a Hubble time), then mergers of the type
discussed here will continue to be identified as cooling flows on the
basis of their cooling times.

The work presented here is only preliminary in nature since significant
parameter space is left to be explored. We have not addressed details of
the gas and dark matter core structure, both of which could
be important. Recent gravitational lensing experiments indicate
that central dark matter distributions may be considerably more
concentrated (i.e. cuspy) than depicted here. Furthermore, cosmological
numerical simulations of the evolution of dark matter halos support the
steep models for the dark matter (e.g., Navarro, Frenk, \& White 1997). 
In general, we believe that a steeper profile could enhance the survivability 
of the cooling flow region during the merger due to the deeper gravitational 
potential and/or favor a rapid re-start of a disrupted cooling flow. 
On the other hand, the deep gravitational potentials generated by these 
steep profiles could also lead to more violent head-on mergers. In this 
respect, the degree of disruption may be enhanced. However, another difference between a NFW type of profile and our models is 
that in these steeper models the baryon fraction increases as a 
function of radius. Since the subcluster will encounter a larger 
amount of gas during its fall, it is very likely that it will not
penetrate as deep as in our models and stop further from the
cooling flow center. Another interesting case not addressed in our study is the
merger of two cooling flow clusters. As our simulations show, any process
or situation that increases the ram pressure of the infalling clusters will
have a disruptive effect on the cooling flows. Thus, the merger of two
cooling flows will probably disrupt both cooling flows.
We have also not addressed off-axis mergers (Roettiger et al. 1998; Ricker 1998). Cooling 
flow survivability could be greatly enhanced if the merger is only 
marginally off-axis. Finally, there are always potential resolution 
effects. To this end, future work will employ higher resolution, fully
3-dimensional simulations which will allow off-axis merger, cuspier
dark matter distributions and a more detailed look at the roll
of post-merger turbulence in cooling flow disruption.

\acknowledgments

This work was partially supported by a NASA Long Term Space Astrophysics 
grant NAGW-3152 and NSF 
grants AST93-17596 and AST98-96039 to JOB. PLG would like to 
acknowledge additional support received by a NASA grant NAG5-3432 and CL and PLG also
acknowledge the support from NASA grant NAG5-3842. We would like to thank 
the Pittsburg Supercomputer Center
TAC committee for granting time to perform our simulations on the C90 and the College of Engineering at New Mexico State University for access to their Cray YMP-EL, W. Gibbs for guiding discussions related to the Poisson equation, Anatoly Klypin for useful discussions about N-body, shock, and entropy evolution, C. Sarazin for insightful comments regarding the project, and an anonymous referee for 
very detailed and lucid comments and suggestions. ZEUS-3D was developed and maintained
by the Laboratory of Computational Astrophysics (LCA) at the National Center for Supercomputing
Applications (NCSA).

\newpage

\clearpage

\figcaption[f1.ps]{Radial gas density (top) and temperature profiles (bottom) for the 400M$_\odot$/year (asterisks) and the 100M$_\odot$/year (diamonds) steady-state cooling flows.\label{Figure 1}}

\figcaption[f2.ps]{Plot of the evolution of the gravitational potential minimum as a function of time for the 1:4 (solid line) and 1:16 (dashed line) mass ratio mergers. The time is relative to the time of core crossing. Note that the most dramatic increase (absolute) in the gravitational potential lasts for about 1 Gyr. \label{Figure 2}}

\figcaption[f3.ps]{Contours of the logarithm of the gas density for run \#8. The grey scale maps represent the logarithm of the density of the passive scalar that traces the subcluster gas. Note that the times are relative to the time of core crossing and that the axes are labeled in units of Mpc. The same contour levels were applied to all the 6 panels in this figure. \label{Figure 3}}

\figcaption[f4.ps]{Plot of the gas density along the merger axis for the merger \#8 (left) and merger \#7 (right). The epochs correspond to (from top to bottom) 0.25, 1.0, and 5.5 Gyrs.
The shaded plot corresponds to the secondary cluster passive scalar. Furthermore, the vertical dotted line shows the location of the gravitational potential minimum in each epoch. Note that for the same epoch, the secondary
gas penetrates deeper and in larger quantities in merger \#8 than in merger \#7. The x axis is labeled in units of 20 kpc. The time is relative to the moment of core crossing.
\label{Figure 4}}

\figcaption[f5.ps]{Contours of the logarithm of the gas density for run \#7. The grey scale maps represent the logarithm of the density of the passive scalar that traces the subcluster gas. Note that the times are relative to the time of core crossing and that the axes are labeled in units of Mpc. The same contour levels were applied to all the 6 panels in this figure. \label{Figure 5}}

\figcaption[f6.ps]{Color plots of the gas temperature distribution at different epochs during the merger for the simulation \#8 (top) and \#7 (bottom). The merger epochs correspond to (from left to right) 0.25, 1.0, and 5.5 Gyrs relative to the time of core crossing. The cooling flow is the dark region located at the core of the primary cluster. All of these panels have the same color scale that shows the hottest regions in white. The axes are labeled in units of Mpc. \label{Figure 6}}

\figcaption[f7.ps]{Plot of the evolution of the primary cluster cooling time as a function of time for the 1:4 mass ratio mergers (bottom) and 1:16 mass ratio mergers (top). The different lines represent the different mergers with different cooling flow strengths and baryon mass fraction as indicated. The
numbers in between parenthesis correspond to the labels in Table 1.
The time is relative to the moment of core crossing.\label{Figure 7}}

\figcaption[f8.ps]{Plot of the evolution of the primary cluster temperature as a function of time for the 1:4 mass ratio mergers (bottom) and 1:16 mas ratio mergers (top). The different lines represent the different mergers with different cooling flow strengths and baryon mass fraction as indicated. The
numbers in between parenthesis correspond to the labels in Table 1.
The time is relative to the moment of core crossing.\label{Figure 8}}

\figcaption[f9.ps]{Plot of $v_{s}$ and $v_{bal}$ as a function of radius for the 1:4 mass ratio mergers (right) and 1:16 mass ratio mergers (left). The symbols represent the different types of mergers (see Table 1). The solid line with no symbols in each panel represents the $v_{ram}$ as computed from the relative velocities of the N-body particles belonging to each cluster.
\label{Figure 9}}

\clearpage
\begin{deluxetable}{c c c c c c c c c c l}
\tablenum{1}
\tablecolumns{10}
\tablewidth{0pt}
\tablecaption{Simulation Data}
\tablehead{
\colhead{label}  &	
\colhead{mass} & 
\multicolumn {2}{c}{\#  of}  &
\multicolumn {2}{c}{$r_c$}         & 
\colhead{cooling}    &
\colhead{central}  &
 \colhead{baryon}  &
\colhead{T} &
\colhead{level of}\\
\colhead{} & 
\colhead{ratio} & 
\multicolumn {2}{c}{particles} &
\multicolumn {2}{c}{(kpc)} &
\colhead{flow} &
\colhead{density} &
\colhead{fraction} &
\colhead{(keV)} &
\colhead{disruption}\\
\colhead{} &
\colhead{} &
\colhead{} &
\colhead{} &
\colhead{} &
\colhead{} &
\colhead{(M$_\odot$/year)} &
\colhead{(cm$^{-3}$)} &
\colhead{\%} &
\colhead{} &
\colhead{of CF}\\
\colhead{} &
\colhead{} &
\colhead{P} &
\colhead{S} &
\colhead{P} &
\colhead{S} &
\colhead{P} &
\colhead{S} &
\colhead{S} &
\colhead{S}&
\colhead{}}
\startdata
1	& 1:4 	& 30000 & 7500 	& 250 		& 157	& 100	& 0.0015	& 15.0	& 6.6	& strong 	 \nl
2	& 1:4 	& 30000 & 7500  & 250 		& 157	& 100	& 0.0006	& 5.0	& 6.6	& mild	 	 \nl
3	& 1:4 	& 30000 & 7500 	& 250 		& 157	& 100	& 0.0003	& 2.5	& 6.6	& mild		 \nl
4	& 1:4 	& 30000 & 7500 	& 250 		& 157	& 400	& 0.0006	& 5.0	& 6.6	& strong		 \nl
5	& 1:16 	& 30000 & 1875 	& 250 		& 99	& 100	& 0.0006	& 5.0	& 2.6	& mild		 \nl
6	& 1:16 	& 30000 & 1875 	& 250 		& 99	& 100	& 0.0012	& 10.0	& 2.6	& strong		 \nl
7	& 1:16 	& 30000 & 1875 	& 250 		& 99	& 400	& 0.0006	& 5.0	& 2.6	& none		 \nl
8	& 1:16 	& 30000 & 1875 	& 250 		& 99	& 400	& 0.0012	& 10.0	& 2.6	& strong	 \nl
9  	& 1:16 	& 30000 & 1875 	& 250 		& 99	& 400	& 0.0001	& 1.2	& 2.6	& none		 \nl
\tablecomments{P and S refer to the values in the primary and secondary cluster respectively.}
\enddata
\end{deluxetable}
\clearpage

\begin{figure}
\plotfiddle{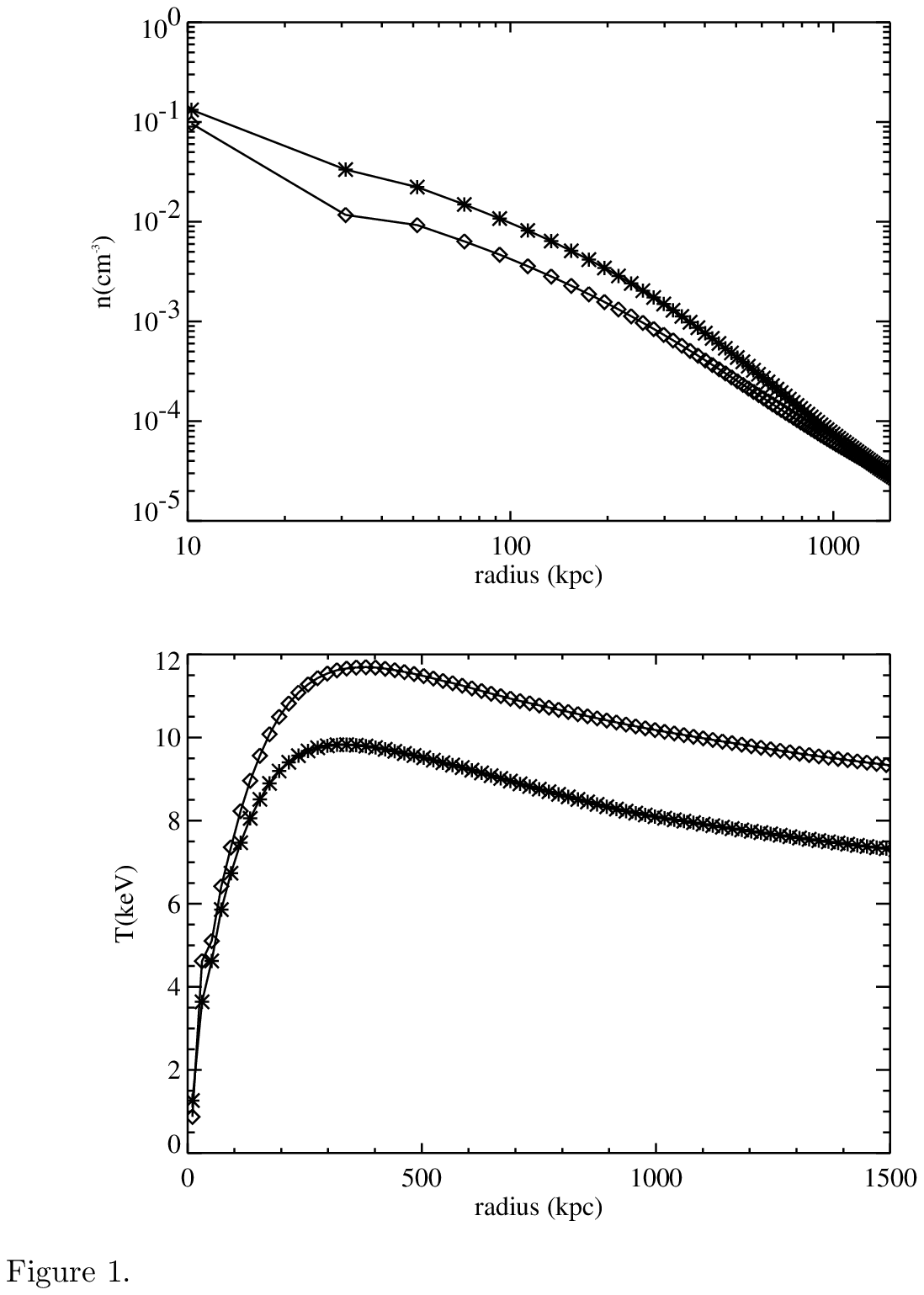}{4in}{0}{100}{100}{-100}{-300}
\end{figure}

\clearpage
\begin{figure}
\plotfiddle{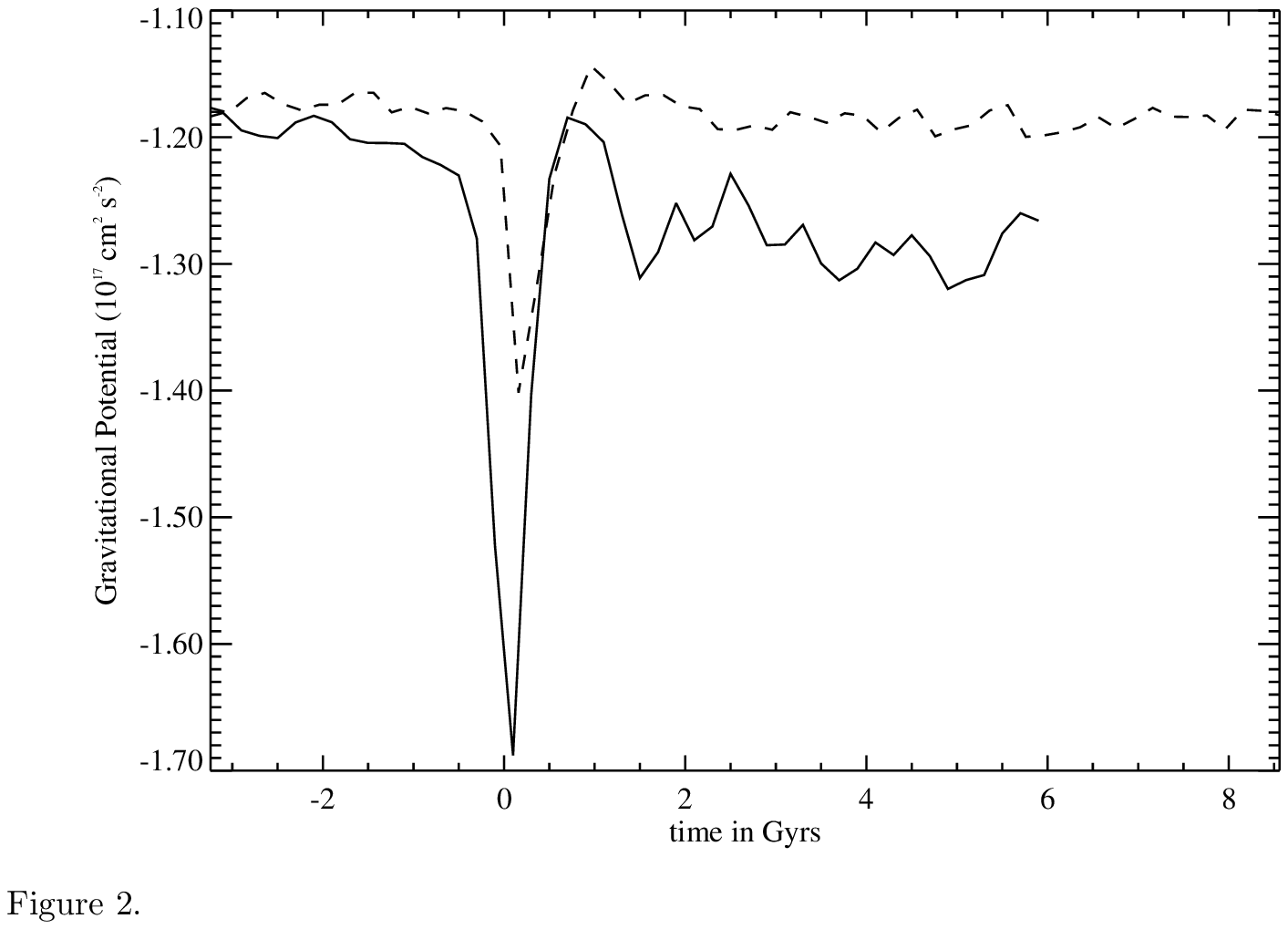}{4in}{0}{100}{100}{-150}{-300}
\end{figure}

\clearpage
\begin{figure}
\plotfiddle{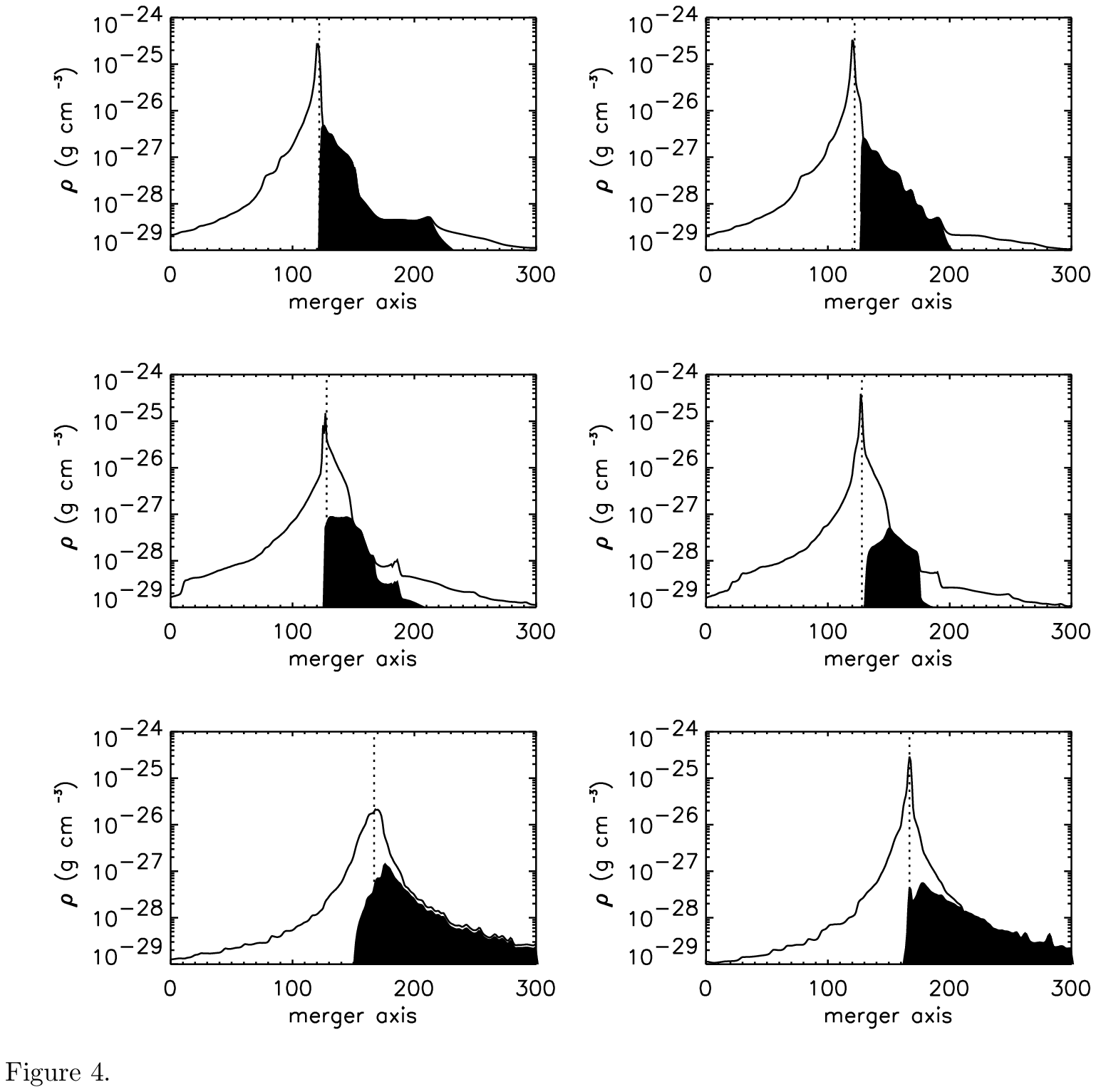}{4in}{0}{100}{100}{-100}{-300}
\end{figure}

\clearpage
\begin{figure}
\plotfiddle{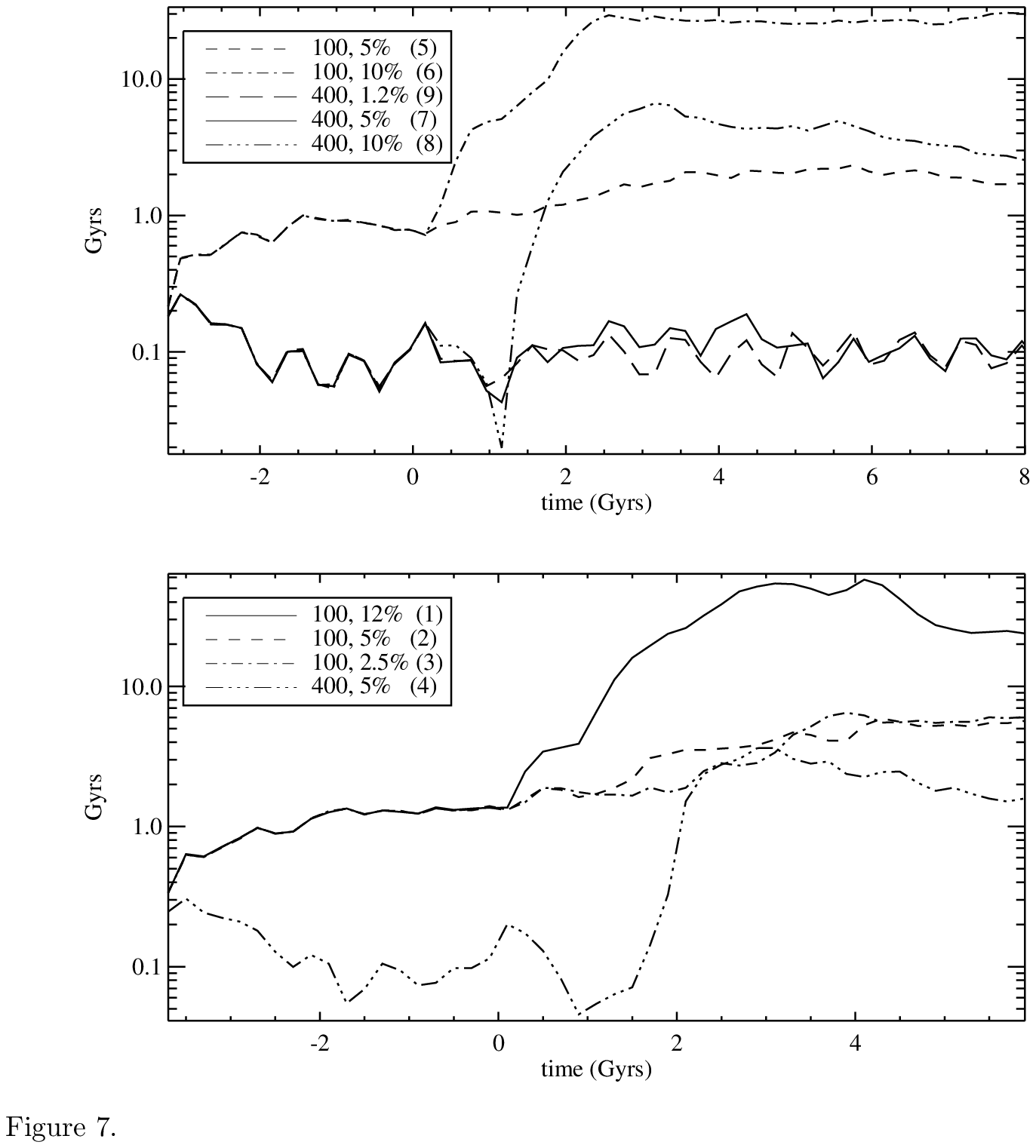}{3.5in}{0}{100}{100}{-100}{-300}
\end{figure}
\clearpage

\begin{figure}
\plotfiddle{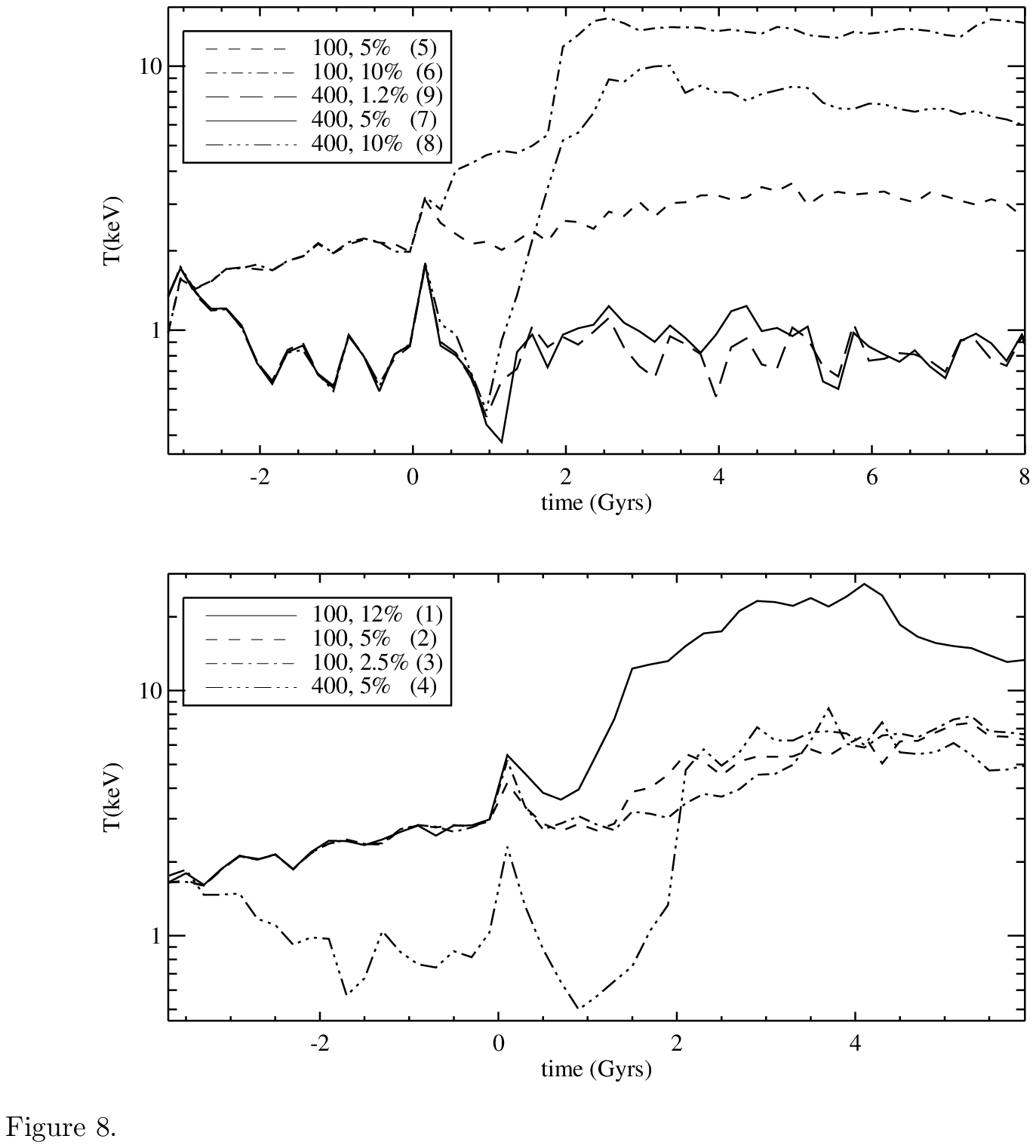}{4in}{0}{100}{100}{-100}{-300}
\end{figure}
\clearpage

\begin{figure}
\plotfiddle{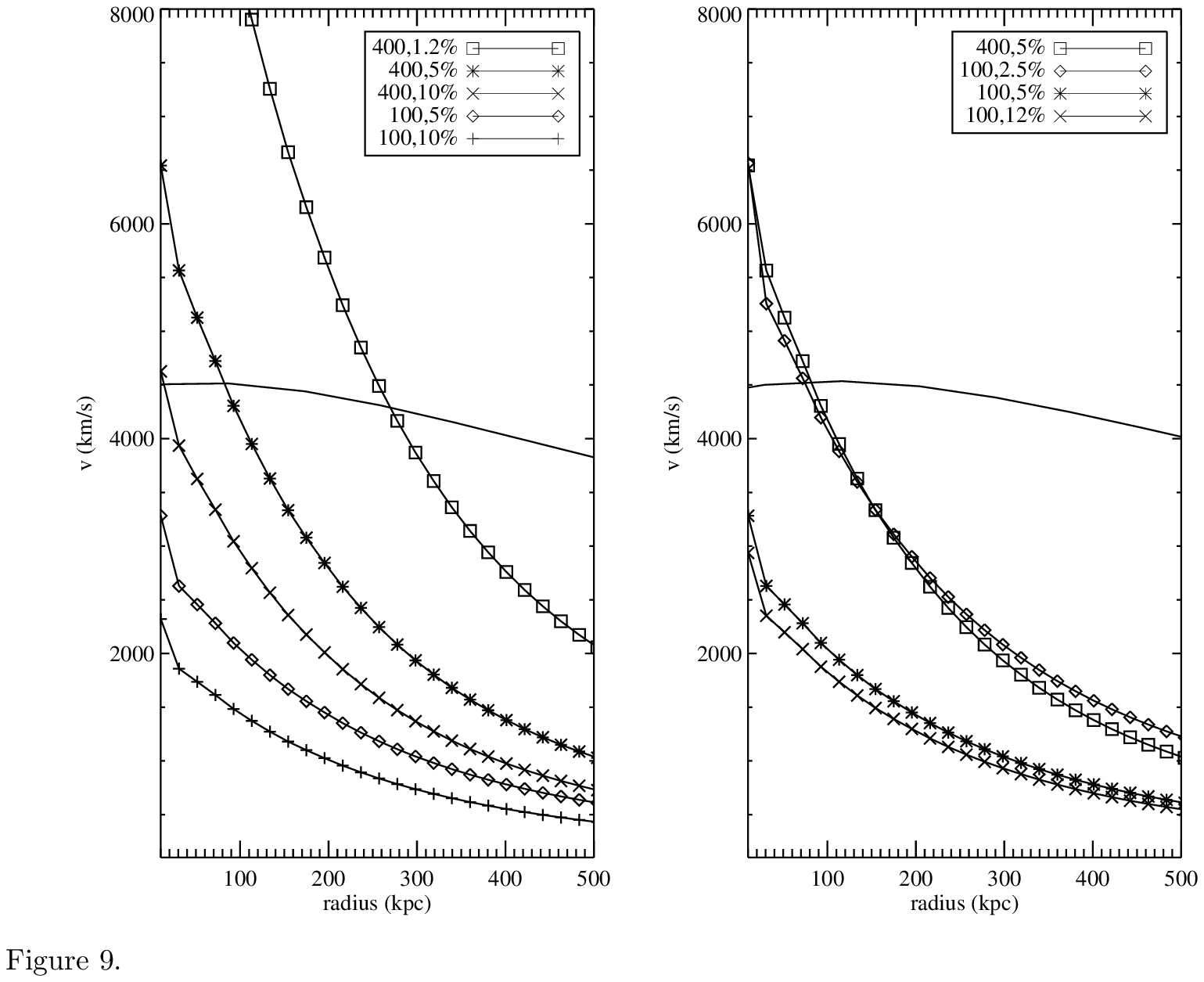}{4in}{0}{100}{100}{-150}{-300}
\end{figure}

\end{document}